\documentclass[
reprint,
superscriptaddress,
showpacs,
preprintnumbers,
nofootinbib,
nobibnotes,
amsmath,
amssymb, 
aps,
prd,
floatfix
]{revtex4-1}
\pdfoutput=1 
\usepackage[utf8]{inputenc} 
\usepackage{soul}
\usepackage{graphicx}  
\usepackage{dcolumn}   
\usepackage{bm,relsize}        
\usepackage{amssymb,amsmath,mathrsfs}
\usepackage{textcomp}
\usepackage{wasysym}
\usepackage{slashed}
\usepackage{multirow}
\usepackage{lipsum, color}
\usepackage[colorlinks=true,allcolors=purple]{hyperref}
\usepackage[usenames,dvipsnames,svgnames]{xcolor}
\usepackage{booktabs}
\usepackage{xfrac}
\usepackage[capitalize]{cleveref}
\usepackage{isotope}
\usepackage{accents}
\usepackage{comment}
\usepackage{physics}
\usepackage{titlesec}
\usepackage{todonotes}
\usepackage{orcidlink}
\usepackage[normalem]{ulem}

\def\ltap{\ \raise.3ex\hbox{$<$\kern-.75em\lower1ex\hbox{$\sim$}}\ }
\def\gtap{\ \raise.3ex\hbox{$>$\kern-.75em\lower1ex\hbox{$\sim$}}\ }
\def\lsim{\ \raise.3ex\hbox{$<$\kern-.75em\lower1ex\hbox{$\sim$}}\ }
\def\gsim{\ \raise.3ex\hbox{$>$\kern-.75em\lower1ex\hbox{$\sim$}}\ }
\def\beq{\begin{equation}}
\def\eeq{\end{equation}}
 \def\be{\begin{equation}} \def\ee{\end{equation}}
\def\bea{\begin{eqnarray}} \def\eea{\end{eqnarray}}

\usepackage{fontawesome5} 
\definecolor{blue-violet}{rgb}{0.33, 0.17, 0.89}

\titlespacing{\subsection}{0pt}{0.1in}{0.1in}

\begin{document}
\preprint{FERMILAB-PUB-26-0111-T}

\title{Impact of New Physics on the JUNO–Long-Baseline Synergy in Neutrino Mass Ordering Determination}

\author{Gustavo F. S. Alves\orcidlink{0000-0002-8269-5365}}
\email{gustavo.alves@northwestern.edu}
\affiliation{Theoretical Physics Department, Fermilab, P.O. Box 500, Batavia, IL 60510, USA}
\affiliation{Northwestern~University,~Evanston,~IL~60208,~USA}
\affiliation{Instituto de F\'isica, Universidade de S\~ao Paulo, C.P. 66.318, 05315-970 S\~ao Paulo, Brazil}

\author{Hiroshi~Nunokawa\orcidlink{0000-0002-3369-0840}}
\email{nunokawa@puc-rio.br}
\affiliation{Departamento de F\'isica, Pontif\'icia Universidade Cat\'olica do Rio de Janeiro,\\
Rua Marquês de São Vicente 225, Rio de Janeiro, Brazil}

\author{Renata Zukanovich Funchal\orcidlink{0000-0001-6749-0022}}
\email{zukanov@if.usp.br}
\affiliation{Instituto de F\'isica, Universidade de S\~ao Paulo, C.P. 66.318, 05315-970 S\~ao Paulo, Brazil}

\date{\today}

\begin{abstract}
The determination of the neutrino mass ordering is one of the flagship goals in particle physics. A well-known and powerful synergy emerges when combining high-precision measurements of the effective atmospheric mass-squared splitting from electron antineutrino disappearance in reactor experiments with that from muon (anti)neutrino disappearance in  accelerator-based long-baseline experiments. To fully exploit this synergy, percent-level precision in the atmospheric mass splitting is required—a target that JUNO is expected to achieve within a few months of data taking. This motivated the formulation of a mass ordering sum rule for neutrino disappearance channels, which shows that by combining data from T2K and NOvA with JUNO after one year of operation, the neutrino mass ordering can be determined at the $3\sigma$ confidence level. Since JUNO has recently started taking data, it is timely to ask whether this sum rule remains robust in the presence of new physics. We identify the necessary conditions for new physics to affect the sum rule and demonstrate that, in some cases, such effects could lead to an incorrect inference of the mass ordering. As concrete examples, we consider Scalar Non-Standard Interactions (SNSI) and neutrinos coupled to an ultralight scalar field. We find that, for SNSI, current constraints render any modification of the sum rule negligible, whereas in the latter case, the inference of the ordering requires caution. Nevertheless, these effects can be disentangled, illustrating how the sum rule can also be used to search for new physics.
\end{abstract}

\maketitle

\section{Introduction} 

Over twenty-five years after the groundbreaking discovery of neutrino flavor oscillations  in 1998~\cite{Super-Kamiokande:1998kpq}, one of the  open fundamental questions in particle physics remains unanswered: the ordering of neutrino masses. If we order the neutrino mass eigenstates by their electron neutrino content -- that is, by identifying $\nu_1$ ($\nu_3$) as the neutrino mass eigenstate with the largest (least)  electron neutrino fraction -- we can formulate the mass ordering problem as equivalent to establishing whether the mass eigenvalues follow the sequence $m_1<m_2<m_3$ (normal ordering) or $m_3<m_1<m_2$ (inverted ordering). We remind the reader that  SNO~\cite{SNO:2011hxd} has determined that $m_2>m_1$ using solar neutrino data. As the field looks to the next generation of precision experiments -- 
JUNO~\cite{JUNO:2015zny,JUNO:2021vlw} in China, which has recently started taking data in August of 2025~\cite{JUNO:2025fpc} and already reported its first physics results~\cite{JUNO:2025gmd},
to upcoming DUNE~\cite{DUNE:2020jqi} in the U.S. and Hyper-Kamiokande~\cite{Hyper-Kamiokande:2018ofw} in Japan, we hope that this enduring enigma will finally be solved in the forthcoming years.

Recently, it has been demonstrated that, thanks to the current one per cent level precision in the determination of the absolute value of the effective atmospheric mass-squared difference~\cite{Nunokawa:2005nx}, $\vert \Delta m^2_{\mu \mu}\vert$~\footnote{We recall for the reader that $\Delta m^2_{ij} \equiv m_i^2 -m_j^2$, for $i,j=1,2,3$. The effective atmospheric $\Delta m^2_{\rm ee} \equiv \cos^2 \theta_{12} \Delta m^2_{31}+ \sin^2\theta_{12}\Delta m^2_{32}$ for $\nu_e$-disappearance and $\Delta m^2_{\rm \mu \mu} \equiv 
\sin^2 \theta_{12} \Delta m^2_{31}+ \cos^2\theta_{12}\Delta m^2_{32} + \sin \theta_{13} \sin 2\theta_{12} \tan \theta_{23} \cos \delta \, \Delta m^2_{21}$ for $\nu_\mu$-disappearance experiments 
were defined for the first time in Ref.~\cite{Nunokawa:2005nx}.
} (or equivalently, $\vert\Delta m_{31}^2\vert$ or $\vert\Delta m_{32}^2\vert$),  achieved by the accelerator-based $\nu_\mu$ disappearance experiments T2K~\cite{T2K:2023mcm} and NOvA~\cite{NOvA:2023iam}, combined with the unprecedented per mill level accuracy expected from JUNO in measuring the magnitude of the effective atmospheric mass-squared difference from $\nu_e$ disappearance, $\vert \Delta m^2_{\rm ee}\vert$ (or equivalently, $\vert\Delta m_{31}^2\vert$ or $\vert\Delta m_{32}^2\vert$), the long-sought determination of the mass ordering may soon be  within reach~\cite{Parke:2024xre}.
Remarkably, this could be achieved through the combination of these complementary (near-) vacuum oscillation measurements alone. This is a direct result of the synergy first identified in Ref.~\cite{Nunokawa:2005nx} and 
further explored in \cite{Li:2013zyd,Cabrera:2020ksc} and 
also recognized as an opportunity by the JUNO collaboration \cite{JUNO:2015zny}.
 In Ref.\cite{Cabrera:2020ksc} this interplay is described as a {\em boost} to the sensitivity of the mass ordering determination.

Determining the neutrino mass ordering through a genuine three-generation vacuum effect mainly through disappearance channels can be viewed as an alternative approach to the more conventional proposals~\cite{Dighe:1999bi,Minakata:2000rx,Minakata:2001qm,Lunardini:2001pb,Barger:2001yr,Barger:2002rr,Petcov:2001sy,Huber:2002rs,Dighe:2003jg,Dighe:2003be,
Lunardini:2003eh,Mena:2004sa,Barger:2005it} that rely on matter effects. However, it also potentially offers a cleaner and more robust pathway -- one in which the existence of beyond the Standard Model interactions would be unlikely to obscure the interpretation. In fact, even for the baselines of T2K (295~km) and NOvA (810~km), it has been shown in Ref.~\cite{Denton:2024thm} that standard matter potential effects to the $\nu_\mu/\bar \nu_\mu$ disappearance channels are at the tenths of the one percent level, while they have been estimated to be even smaller for the $\bar \nu_e$ disappearance channel at the medium baseline (52.5 km) JUNO experiment~\cite{Li:2016txk,Khan:2019doq}. 

Moreover, non-standard matter potentials resulting from  the interactions of neutrinos with matter, mediated by a vector particle, due to the stringent constraints imposed by neutrino oscillation and coherent  elastic neutrino-nucleus scattering data~\cite{Coloma:2023ixt},  have been shown to affect only slightly JUNO's precision~\cite{Martinez-Mirave:2021cvh}.
Therefore, it seems that the determination of 
$\vert \Delta m^2_{\rm e e}\vert$ and $\vert \Delta m^2_{\rm \mu \mu}\vert$ by these disappearance experiments could be safely considered to occur effectively in vacuum.

In this work, we revisit the neutrino mass ordering and the associated sum rule~\cite{Parke:2024xre}, examining how it may be affected by the presence of new  physics. In Section~\ref{sec:nmo}, we begin with a general discussion of how the mass-ordering sum rule can be modified. Then, in Section~\ref{sec:exp}, we demonstrate that the JUNO–long-baseline synergy could, under certain conditions, lead to an incorrect conclusion about the neutrino mass ordering, and we quantify the size of the modifications required for this to occur.

In Sections~\ref{sec:SNSI} and~\ref{sec:ULDM}, we explore two specific examples of beyond-the-Standard-Model effects that induce such modifications: scalar non-standard interactions (SNSI), as proposed in~\cite{Ge:2018uhz}, and an ultralight scalar field, studied, e.g. in~\cite{Krnjaic:2017zlz,Dev:2020kgz,Losada:2021bxx,Dev:2022bae,Berlin:2016woy,Cordero:2022fwb,Delgadillo:2025wxw}. Finally, in Section~\ref{sec:conc}, we summarize our main findings and present concluding remarks.

\section{How new physics may affect the mass ordering sum rule}
\label{sec:nmo}
The mass ordering sum rule introduced in Ref.~\cite{Parke:2024xre} clearly demonstrates why  ordering sensitivity is enhanced by combining disappearance data  from T2K/NOvA (LBL) and reactor (R) experiments. This approach exploits the fact that while  
disappearance experiments before JUNO lack individual sensitivity to the mass ordering, this landscape will be transformed by the newly started JUNO experiment.
For JUNO (JU), as it was shown in Ref.~\cite{Forero:2021lax}, the expected best fit value of the atmospheric mass squared difference for normal ordering (NO) and inverted ordering (IO) will differ approximately by $1.8 \times 10^{-5}~\rm{eV}^2$, bringing the mass ordering sum rule to the form
\begin{widetext}
    \begin{align}
        ( \Delta m^2_{31}\vert^{\rm NO}_{\rm LBL} - \Delta m^2_{31}\vert^{\rm NO}_{\rm JU}) + (\vert \Delta m^2_{32}\vert^{\rm IO}_{\rm JU} - \vert \Delta m^2_{32}\vert^{\rm IO}_{\rm LBL}) \approx (2 \cos 2 \theta_{12}- 2 \sin \theta_{13}  \overline{\cos \delta} )\, \Delta m^2_{21} + 1.8 \times 10^{-5} \rm eV^2\, ,
        \label{eq:sum_rule}
    \end{align}
\end{widetext}
where $\Delta m^2_{31}\vert^{\rm NO}_{\rm exp}$ ($\vert \Delta m^2_{32}\vert^{\rm IO}_{\rm exp}$)
is the best fit value assuming NO (IO) using exp=LBL, JU data, $\overline{\cos \delta}$ is the average value  
of $\cos \delta$  for NO and IO fits, with $\delta$ being the CP phase. In Eq.~\eqref{eq:sum_rule} we also have approximated $\sin 2\theta_{12}\tan\theta_{23}\approx 1$. This serves as a guiding principle for understanding how new physics can modify the mass ordering sum rule. In general, new effects could
\begin{enumerate}
    \item Change the best-fit values for NO and IO within a single experiment -- for instance, 
    $\vert \Delta m^2_{\rm ee}\vert^{\rm IO}_{\rm JU}= \Delta m^2_{\rm ee}\vert^{\rm NO}_{\rm JU} +  1.8 \times 10^{-5}\, {\rm{eV^2}}+\delta m^2$. Such a shift ($\delta m^2$) introduces an additional term to the right-hand side of Eq.~\eqref{eq:sum_rule}, analogous to the displacement between the NO and IO fits in JUNO.
    \item Introduce a mass shift, $\delta m^2_{\rm LJ}$, caused by some new physics,  across the different experiments. Assuming that the magnitude of the shift is smaller than $\Delta m^2_{31}|^{\rm NO}$ or $|\Delta m^2_{32}|^{\rm IO}$, this scenario implies $|\Delta m^2_{31}\vert^{\rm NO}_{\rm LBL}-\Delta m^2_{31}\vert^{\rm NO}_{\rm JU} = \delta m^2_{\rm LJ}$ ($|\Delta m^2_{32}|^{\rm IO}_{\rm JU} - |\Delta m^2_{32}|^{\rm IO}_{\rm LBL} = \delta m^2_{\rm LJ}$) for the case where the true ordering is normal (inverted).
\end{enumerate}
As the first case only  change within a single experiment the difference in the mass ordering (similarly to what happens when we include the difference in JUNO's best fit), we focus on the second possibility that provides a richer phenomenology. In this case the sum rule is modified to
\begin{widetext}
    \begin{align}
        ( \Delta m^2_{31}\vert^{\rm NO}_{\rm LBL} - \Delta m^2_{31}\vert^{\rm NO}_{\rm JU}- \delta m^2_{\rm LJ}) + (\vert \Delta m^2_{32}\vert^{\rm IO}_{\rm JU} - \vert \Delta m^2_{32}\vert^{\rm IO}_{\rm LBL}-\delta m^2_{\rm LJ})\approx  (2 \cos 2 \theta_{12}- 2 \sin \theta_{13}  \overline{\cos \delta} )\, \Delta m^2_{21} \nonumber\\  + 1.8 \times 10^{-5} \rm eV^2\,.
        \label{eq:sum_rule_NP}
    \end{align}
\end{widetext}
This  means that if NO is true, then  we must find, by combining LBL with  JUNO, that 
\begin{eqnarray}
    \Delta m^2_{31}\vert^{\rm NO}_{\rm LBL} = \Delta m^2_{31}\vert^{\rm NO}_{\rm JU} + \delta m^2_{\rm LJ}\, ,
\label{eq:J-LBL-NO}
\end{eqnarray}
or instead, if IO is true, that  
\begin{eqnarray}
    \vert\Delta m^2_{32}\vert^{\rm IO}_{\rm LBL} = \vert \Delta m^2_{32}\vert^{\rm IO}_{\rm JU} - \delta m^2_{\rm LJ}\, .
\label{eq:J-LBL-IO}
\end{eqnarray}
This anticipates that even if NO is the true ordering in nature, a non-zero $\delta m_{\rm LJ}^2$ could bias the conclusion by compensating for the fit discrepancies in the IO hypothesis. In the following section, we demonstrate that for specific values of 
$\delta m^2_{\rm LJ}$, the sensitivity is not merely degraded; instead, the analysis can lead to an incorrect inference of the mass ordering.

\section{Combining T2K/NOvA with JUNO with new physics}
\label{sec:exp}

\begin{figure*}[!t]
    \centering
    \begin{minipage}{0.48\textwidth}
        \centering
        \includegraphics[width=\textwidth]{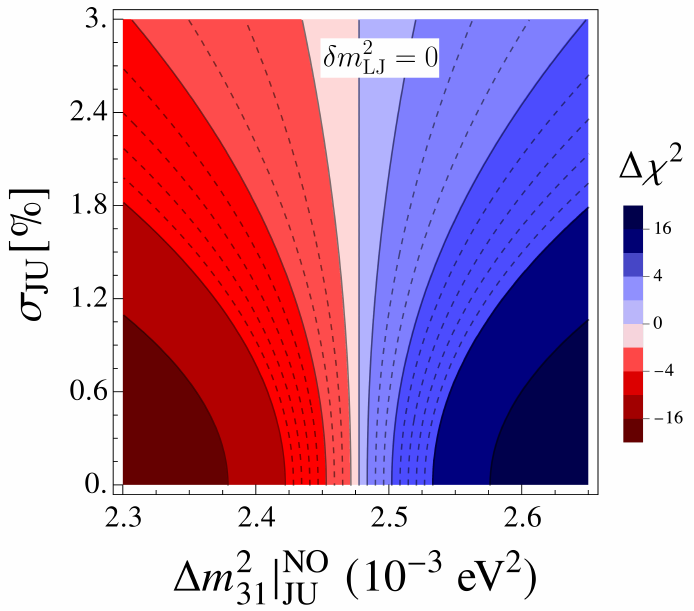}
    \end{minipage}
    \hfill
    \begin{minipage}{0.48\textwidth}
        \centering
        \includegraphics[width=\textwidth]{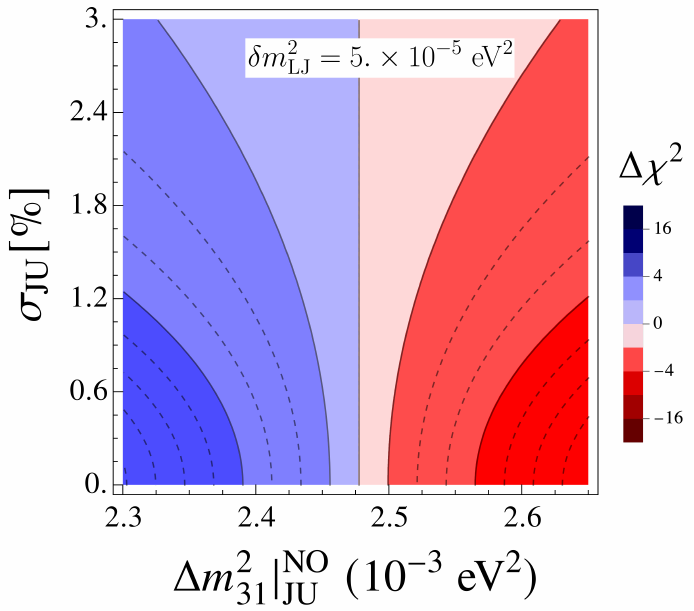}
    \end{minipage}
    \caption{Iso-contours of $\Delta \chi^2(\Delta m^2_{31}\vert^{\rm NO}_{\rm JU},\sigma_{\rm JU}$,$\delta m^2_{\rm LJ})$
    representing the mass ordering preference when combining JUNO and LBL data. The blue 
    region corresponds to values of $\Delta m^2_{31}\vert^{\rm NO}_{\rm JU}$, as measured by JUNO, for 
    which the synergy favors NO, while the red region corresponds to values where IO is 
    favored. The left plot reproduces the standard fan plot of Ref.~\cite{Parke:2024xre}, while 
    the right plot shows a scenario where new physics introduces a displacement between the 
    LBL and JUNO best-fit values. Notably, this displacement is sufficient to incorrectly 
    infer the neutrino mass ordering.}
    \label{fig:fan_plots_NSI}
\end{figure*}

To evaluate the sensitivity to the mass ordering by combining muon and electron neutrino disappearance data in the presence of new physics effects in a simplified  yet quite accurate way,  given the high precision of these experiments, we consider the following $\chi^2$ function 
\begin{equation}
   \chi^2_{\rm exp}(\Delta m^2;\sigma_{\rm exp},\Delta m^2_{\rm atm}) = \left(\frac{\Delta m^2 - \Delta m^2_{\rm atm}}{\sigma_{\rm exp}} \right)^2,
\end{equation}
where $\Delta m^2=\Delta m^2_{31}$ ($\vert \Delta m^2_{32}\vert$) for NO (IO). Here $\Delta m^2_{\rm atm}$ is the best fit value of the experiment under the NO or IO hypothesis, and $\sigma_{\rm exp}$ is the measurement precision. We take the best fit values for the long baseline neutrino experiments $\Delta m^2_{31}|^{\rm NO}_{\rm LBL} = 2.516\times 10^{-3}~\text{eV}^2$, $\vert \Delta m^2_{32}\vert^{\rm IO}_{\rm LBL} = 2.485\times 10^{-3}~\text{eV}^2$, and $\sigma_{\rm LBL} = 0.031\times 10^{-3}~\text{eV}^2$ from Ref.~\cite{NuFIT2022}. 
 In JUNO's case  $\Delta m^2_{31}|^{\rm NO}_{\rm JU}$, $\delta m^2_{\rm LJ}$, as well as, $\sigma_{\rm JU}$ are treated as free variables.
By combining the measurements from LBL and JUNO, we can define
\begin{widetext}
    \begin{align}
        \begin{split}
        \chi^2_{\rm NO}(\Delta m^2_{31};\Delta m^2_{31}\vert^{\rm NO}_{\rm JU}, \sigma_{\rm JU},\delta m^2_{\rm  LJ}) &= \chi^2_{\rm LBL}(\Delta m^2_{31}) + \chi^2_{\rm JU}(\Delta m^2_{31}+\delta m^2_{\rm LJ}; \sigma_{\rm JU},\Delta m^2_{31}\vert^{\rm NO}_{\rm JU}),\\
            \chi^2_{\rm IO}(\vert \Delta m^2_{32}\vert; \Delta m^2_{31}\vert^{\rm NO}_{\rm JU},  \sigma_{\rm JU},\delta m^2_{\rm  LJ}) &= \chi^2_{\rm LBL}(\vert \Delta m^2_{32}\vert) + \chi^2_{\rm JU}(\vert\Delta m^2_{32}\vert -\delta m^2_{\rm LJ}; \sigma_{\rm JU},\vert\Delta m^2_{32}\vert^{\rm IO}_{\rm JU}),\\
    \end{split}
    \label{eq:chi2}
\end{align}
\end{widetext}
where we omit the explicit dependence on the best fit value and precision in the $\chi^2_{\rm LBL}$ as they are fixed. These functions depend on JUNO best fit for NO $\Delta m^2_{31}\vert^{\rm NO}_{\rm JU}$, to be determined by the actual data, the precision of the measurement  $\sigma_{\rm JU}$ and the mass squared shift due to new physics $\delta m^2_{\rm LJ}$. JUNO's best fit for NO and IO are related as
$\vert\Delta m^2_{32}\vert^{\rm IO}_{\rm JU}=\Delta m^2_{31}\vert^{\rm NO}_{\rm JU}+4.7 \times 10^{-5} \; \rm eV^2$~\cite{Parke:2024xre}.  
We  now minimize $\chi^2_{\rm NO}$ and $\chi^2_{\rm IO}$ for fixed values of these parameters to obtain $\chi^2_{\rm NO}\vert^{\rm min}$ and $\chi^2_{\rm IO}\vert^{\rm min}$.

To estimate the sensitivity to a particular neutrino mass ordering for non zero $\delta m_{\rm LJ}^2$ we define
\begin{equation*}
    \Delta \chi^2(\Delta m^2_{31}\vert^{\rm NO}_{\rm JU},\sigma_{\rm JU},\delta m^2_{\rm LJ}) \equiv \chi^2_{\rm IO}\vert^{\rm min} - \chi^2_{\rm NO}\vert^{\rm min},
\end{equation*}
such that, if $\Delta \chi^2 > 0$ ($\Delta \chi^2 < 0$), NO (IO) is favored. 
Explicitly, we have
\begin{widetext}
    \begin{equation}
        \Delta \chi^2
        =   \frac{2 (\Delta m^2_{31}\vert^{\rm NO}_{\rm JU}-\Delta^{*}_{31})
        (\Delta m^2_{31}\vert^{\rm NO}_{\rm LBL}-
        |\Delta m^2_{32}\vert^{\rm IO}_{\rm LBL}+4.7\times 10^{-5}\,\text{eV}^2-2\, \delta m^2_{\rm LJ})}{\sigma_{\rm LBL}^2+\sigma_{\rm JU}^2},
    \label{eq:delchi2snsi}  
    \end{equation}
\end{widetext}
where 
\begin{eqnarray}
     \Delta^{*}_{31} &\equiv&
  \frac{\Delta m^2_{31}\vert^{\rm NO}_{\rm LBL}+\vert \Delta m^2_{32}\vert^{\rm IO}_{\rm LBL}- 4.7 \times 10^{-5} \rm eV^2}         {2}\nonumber \\
        &\approx& 2.477 \times 10^{-3}\, \text{eV}^2.
    \label{eq:Delta_star}
\end{eqnarray}

In Fig.~\ref{fig:fan_plots_NSI}, to illustrate the effect, we present the iso-contours of  $\Delta \chi^2 (\Delta m^2_{31}\vert^{\rm NO}_{\rm JU},\sigma_{\rm JU},\delta m^2_{{\rm LJ}})$ for two cases: $\delta m^2_{\rm LJ}=0$~eV$^2$, i.e., neutrinos have only interactions expected in the Standard Model (SM), and $ \delta m^2_{\rm LJ} = 5 \times 10^{-5}$~eV$^2$, where new physics impacts the synergy of JUNO and LBL experiments. The blue region corresponds to the values of $\Delta m^2_{31}\vert^{\rm NO}_{\rm JU}$ as measured by JUNO that would favor NO when combined with LBL data, while the red region represents values where IO would be favored. The left fan plot reproduces the  standard result of Ref.~\cite{Parke:2024xre}; the right panel shows how the standard plot changes if $\delta m^2_{\rm LJ} = 5 \times 10^{-5}$~eV$^2$. Notably, the mass ordering preference is completely swapped.

It is straightforward to understand this behavior analytically. From Eq.~\eqref{eq:delchi2snsi}, we see that independent of the presence of new physics effects, $\Delta \chi^2 = 0$, {\it i.e.}, NO and IO cannot be distinguished, for $\Delta m^2_{31}\vert^{\rm NO}_{\rm JU}= \Delta^{*}_{31} $, basically fixed by the best fit values of the LBL experiments.
In addition, we can see that $\Delta \chi^2 = 0$ also when 
\begin{eqnarray}
    \delta m^2_{\rm LJ} 
    &= &
   \frac{1}{2} (\Delta m^2_{31}\vert^{\rm NO}_{\rm LBL}-
    |\Delta m^2_{32}\vert^{\rm IO}_{\rm LBL}+4.7\times 10^{-5}\,\text{eV}^2)\nonumber \\
    &\approx& 4 \times 10^{-5}\,\text{eV}^2.
\end{eqnarray}
Hence, the sign of $\Delta \chi^2$ is controlled by two factors: whether 
 $\Delta m^2_{31}\vert^{\rm NO}_{\rm JU}$ is larger or smaller than the critical value $\Delta^{*}_{31}$, 
and whether $\delta m^2_{\rm LJ}$ is larger or smaller than 
$4 \times 10^{-5}$~eV$^2$. This means that a shift in the best-fit values of JUNO 
and LBL of the order of a few percent of $\Delta m^2_{\rm atm}$ can complicate the 
determination of the mass ordering, undermining the expected synergy from the combination 
of LBL $\nu_\mu$ and JUNO $\nu_e$ disappearance data.

Having discussed the possibilities from a general perspective we now turn to concrete examples of how such modifications could happen in the presence of some beyond SM new physics effects.

\section{Scalar nonstandard interactions}
 \label{sec:SNSI}

First as proposed in \cite{Ge:2018uhz} we consider a new, light, scalar mediator $\phi$, with couplings to the neutrinos $\nu$ and matter fields $\psi$. The relevant part of the Lagrangian is
\begin{equation}
    \mathcal{L} \supset - (m_\nu + y_\nu \phi) \overline{\nu}\nu - y_\psi \overline{\psi}\psi
    \label{eq:nu_matter_coupling_to_phi},
\end{equation}
where $y_\nu$ and $y_\phi$ are couplings of the mediator to neutrino and matter fields, respectively,
showing that this scalar will affect the neutrino masses. Assuming that the contribution is small, the effective Hamiltonian is given by 
\begin{equation}
  \mathcal H_{\rm eff}  \approx E_\nu + \frac {\left( M + \delta M \right) \left( M + \delta M \right)^\dagger}{2 E_\nu} + V_{\rm m} \,,
    \label{eq:H_nsi}
\end{equation}
where $E_\nu$ is the neutrino energy, $M$ is the neutrino mass matrix, $\delta M$ denotes the mass correction due to SNSI, and $V_m$ is the standard matter potential given by $V_m= \text{diag} (\,\pm \sqrt{2}G_F n_e, 0,0\,)$ with $G_F$ and $n_e$ being, respectively, the Fermi constant and the electron number density in matter; where the $+$ ($-$) sign is for neutrinos (antineutrinos). We can write the neutrino mass shift for Earth-based detectors as~\cite{Smirnov:2019cae}
\begin{equation}
    \delta M = \frac{y_\nu y_\psi n_\psi}{m_\phi^2}F(r,m_\phi) = \eta F(R_{\rm E},m_\phi),
    \label{eq:delta_M}
\end{equation}
where $n_\psi$ is the matter particles number density, $m_\phi$ the scalar mass, $R_{\rm E}$ is the Earth radius, and $F(R_{\rm E},m_\phi)$ is a form factor that depends on the matter density profile, and $\eta$ is a dimensionless matrix which characterizes the SNSI effect whose magnitude is assumed to be small. In Eq.~\eqref{eq:delta_M} it was assumed a spherically symmetric matter profile.

We  focus on the particularly simple case where  $\delta M = \text{diag} (0,0, \eta_{33})F(R_{\rm E},m_\phi)$. In this case, for $|\eta_{33}| \ll 1$, the atmospheric neutrino mass scales $\Delta m^2_{31}$ and $\Delta m^2_{32}$ shift in matter, up to the first order in  $|\eta_{33}|$, to 
\begin{eqnarray}
    {\Delta m^2_{31}}  & \to &   
    {\Delta m^2_{31}} + 2m_3\eta_{33} F(R_{\rm E},m_\phi), \nonumber \\
    {\Delta m^2_{32}} &\to & 
    {\Delta m^2_{32}} +  2m_3\eta_{33} F(R_{\rm E},m_\phi),
    \label{eq:shift}
\end{eqnarray}
while $\Delta m^2_{21}$ remains unchanged. 

The shift described in Eq.~\eqref{eq:shift} may at first appear to be nothing more than a simple redefinition of the mass splittings. While this interpretation is valid for individual measurements, the situation changes when different experiments are compared. In particular, the distinct matter densities relevant for JUNO and LBL experiments lead to different shifts in their respective best-fit values. In neutrino oscillation analyzes, the matter density along the neutrino trajectory is typically assumed to be constant -- a very good approximation~\cite{Koike:1998hy}. See also a recent work~\cite{Li:2025hye} where the impact of the matter density variation along the neutrino trajectory for JUNO was discussed. However, the numerical value of this effective density differs from one experimental setup to another. One can relate the values determined by the long baseline accelerator  $\nu_\mu$ disappearance experiments NOvA and T2K  (denoted as $\Delta m^2_{\rm atm}|_{\rm LBL}$) with the one that will be measured by the reactor $\nu_e$ disappearance experiment JUNO (denoted as $\Delta m^2_{\rm atm}|_{\rm JU}$) under our SNSI scenario as 
\begin{equation}
    \Delta m^2_{\rm atm}|_{\rm LBL}= \Delta m^2_{\rm atm}|_{\rm JU} + \delta m^2_{\rm LJ},
    \label{eq:shiftLJ}
\end{equation}
where $\Delta m^2_{\rm atm} = \Delta m^2_{31},\ \Delta m^2_{32}$, and 
\begin{equation}
    \delta m^2_{\rm LJ} \equiv 2 m_3  \eta_{33}F(R_{\rm E}, m_\phi)\frac{\rho_{\rm LBL} - \rho_{\rm JU}}{\rho_{\rm LBL}}.
    \label{eq:shiftLJ}
\end{equation}
For simplicity and as a reasonable approximation, we assume that the average matter density for NOvA and T2K experiments is the same, denoted as $\rho_{\rm LBL}$, implying 
that these experiments measure the same value of the atmospheric mass squared difference even with the presence of SNSI.

As discussed earlier, we need $\delta m^2_{\rm LJ}\sim  4\times 10^{-5}~\rm{eV}^2$. For the assumed average matter density of LBL $\rho_{\rm LBL} = 2.8~\text{g/cm}^3$~\cite{NOvA:2021nfi} and JUNO $\rho_{\rm JU} = 2.45~\text{g/cm}^3$~\cite{JUNO:2024jaw} we need $\eta_{33}F(R_E,m_\phi) \sim 10^{-3}\times(0.1~\text{eV}/m_3)$. 
We note that the recent work Ref.~\cite{Choubey:2026jiq}, which studied also the impact of SNSI for JUNO, found a similar requirement. Naively, from Eq.~\eqref{eq:delta_M}, this may be possible for very light scalars. However, when $m_\phi < R^{-1}_{\rm E}$ the scaling of $\delta M$ is modified by the form factor~\cite{Smirnov:2019cae, Babu:2019iml} and the correction becomes independent of the scalar mass, such that we must adjust the couplings to reach the required mass shift. Bounds from fifth force searches~\cite{Adelberger:2006dh}, tests of the equivalence principle~\cite{Schlamminger:2007ht} and cosmology~\cite{Kreisch:2019yzn,Venzor:2020ova} severely constrain this possibility. In particular, Ref.~\cite{Smirnov:2019cae} shows that $\delta M \sim 10^{-3}$ is already excluded and only negligible shifts ($\delta M \sim 10^{-11}~\rm{eV}$) are allowed. Therefore, the mass ordering sum rule should not be impacted by SNSI.

\begin{figure*}[t]
    \centering
    \includegraphics[width=0.75\textwidth]{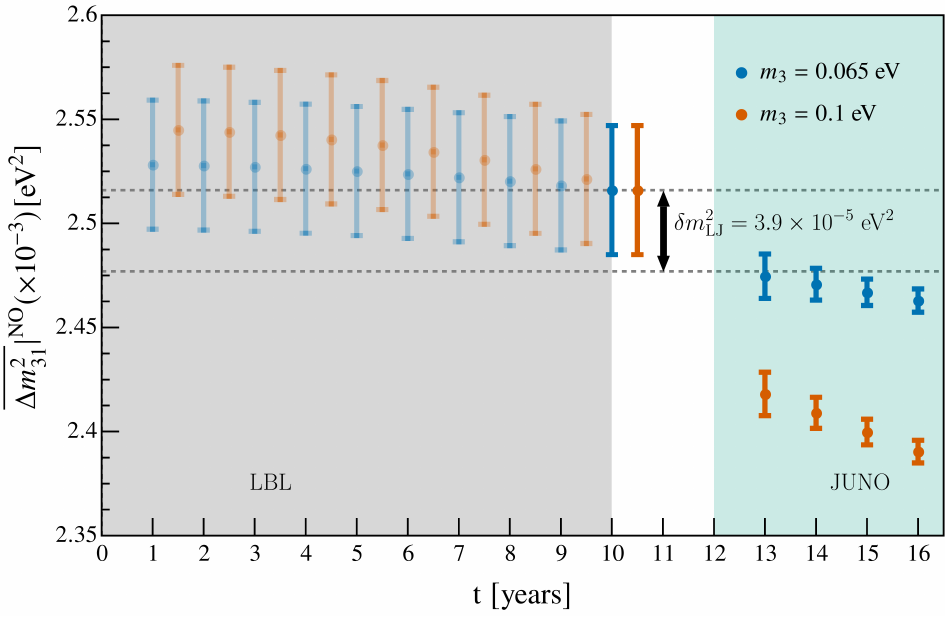}
    \caption{Expected time evolution of the atmospheric mass-splitting best-fit values for LBL experiments and JUNO, shown for two different neutrino mass values: $m_3 = 0.065$~eV (blue) and $m_3 = 0.1$~eV (orange). The scalar field parameters are fixed at $\eta_\phi = 0.01$, $\theta = 0$, and $m_\phi = 2\times10^{-24}$~eV. Time $t=0$ corresponds to the start of LBL data taking, with JUNO beginning operations 12 years later. The vertical arrow indicates the critical shift $\delta m^2_{\rm LJ}$ that could lead to an incorrect mass-ordering inference. LBL uncertainty bars are back-propagated from current measurements, and the points are slightly displaced for visibility. JUNO uncertainty bars follow the expected precision improvement with accumulated statistics. Normal ordering is assumed.}
    \label{fig:best_fit_time_dep}
\end{figure*}
\section{Time dependent neutrino masses}
\label{sec:ULDM}

Next, we examine the scenario where the ultralight scalar field can be treated as a classical stochastic background that modulates the mass of the third neutrino mass eigenstate~\cite{Krnjaic:2017zlz, Dev:2020kgz, Dev:2022bae, Berlin:2016woy}. In this framework, the neutrino mass varies in time as
\begin{equation}
    m_3(t) = m_3 + y_\nu \frac{\sqrt{2\rho_\phi}}{m_\phi}\cos(m_\phi t + \theta),
\end{equation}
where $y_\nu$ is the coupling between neutrino and the scalar field, $\rho_\phi$ is the local energy density of the scalar field, $\theta$ is a phase related to the field initial conditions and $m_\phi$ is the scalar mass. For the discussion that follows we set $\theta =0$ for simplicity. In Appendix~\ref{app:scalar_phase} we comment on the effect of varying the phase. This temporal modulation induces a time dependence in the atmospheric mass splitting. To leading order we find
\begin{equation}
    \Delta m_{31}^2(t)
    = \Delta m_{31}^2 + 2 m_3^2 \eta_\phi \cos(m_\phi t),
\end{equation}
where we define the dimensionless parameter
\begin{equation}
\eta_\phi \equiv \frac{y_\nu \sqrt{2\rho_\phi}}{m_\phi m_3}.
\end{equation}
We assume that the scalar field oscillates slowly enough such that the LBL and JUNO 
measurements of the atmospheric mass splitting are well approximated by their 
time-averaged values over the respective data-taking periods. Specifically, the 
experimental best fit after $\Delta t$ years of data collection corresponds to
\begin{align}
        \overline{\Delta m_{31}^2}  \simeq \Delta m_{31}^2 + 2 m_3^2 \eta_\phi  \frac{\sin(m_\phi \Delta t)}{m_\phi \Delta t}\, .
        \label{eq:avg_LBL}
\end{align}
LBL experiments have collected data for roughly ten years~\cite{2817608, Giganti2024_T2Kstatus}. 
We adopt this timescale for our analysis, as using the exact or approximated exposure periods do not affect our conclusions. 
Therefore, their measured mass splitting should be interpreted as the average over 
$\Delta t_{10}=10~{\rm yrs}$. We take the most recent LBL best fit (assuming normal ordering) as this ten-year average in Eq.~\eqref{eq:avg_LBL}. Solving for the mass splitting $\Delta m^2_{31}$, which corresponds to the actual atmospheric mass splitting only in the absence of the scalar field, i.e., $y_\nu = 0$, we can rewrite the averaged value solely in terms of the LBL best fit and scalar parameters as
\begin{widetext}
    \begin{equation}
         \overline{\Delta m_{31}^2}\big|^{\rm NO}_{\rm LBL}
         =
         \Delta m_{31}^2\big|^{\rm NO}_{\rm LBL}
         + 2 m_3^2 \eta_\phi
         \left[
            \frac{\sin(m_\phi \Delta t_{\rm LBL})}{m_\phi \Delta t_{\rm LBL}}
            - \frac{\sin(m_\phi \Delta t_{10})}{m_\phi \Delta t_{10}}
         \right],
    \end{equation}
\end{widetext}
where $\Delta t_{\rm LBL}$ is the duration over which the LBL measurement is averaged. 
Note that by construction, when $\Delta t_{\rm LBL} = \Delta t_{10}$, we have 
$\overline{\Delta m_{31}^2}\big|^{\rm NO}_{\rm LBL}= \Delta m^2_{31}\big|_{\rm LBL}^{\rm NO}$.

JUNO, however, began taking data at a later time~\cite{JUNO:2025fpc}, offset from the start of LBL operations by an interval $t_{\rm off}\sim 10\text{--}15~{\rm yr}$. During JUNO's run, the ultralight scalar field is at a different phase of its oscillation cycle. This leads to a predicted JUNO measurement of
$\Delta m^2_{31}$ value different from that of LBL experiments even for the true mass ordering as
\begin{widetext}
    \begin{equation}
        \overline{\Delta m_{31}^2}\big|^{\rm NO}_{\rm JU}(\Delta t_{\rm JU}, t_{\rm off})
        =
        \Delta m_{31}^2\big|^{\rm NO}_{\rm LBL}
        - 2 m_3^2 \eta_\phi
          \left[
             \frac{\sin(m_\phi \Delta t_{10})}{m_\phi \Delta t_{10}}
          \right]
        + 2 m_3^2 \eta_\phi
          \left[
             \frac{\sin\!\left(m_\phi (\Delta t_{\rm JU}+t_{\rm off})\right)}{m_\phi \Delta t_{\rm JU}}
             -
             \frac{\sin(m_\phi t_{\rm off})}{m_\phi \Delta t_{\rm JU}}
          \right],
    \end{equation}
\end{widetext}
where $\Delta t_{\rm JU}$ is JUNO's the time scale over which its measurement is averaged over.

Fig.~\ref{fig:best_fit_time_dep} illustrates the typical time evolution of the atmospheric mass splitting best-fit values for both LBL and JUNO experiments, with $t=0$ corresponding to the start of LBL data collection. For definiteness, let us assume that JUNO begins operations 12 years later after the reference initial time considered for LBL $(t=0)$ and reports its first best fit after one year of data taking. The plot shows results for two different neutrino mass values, $m_3 = 0.065$ and $0.1$~eV, to emphasize the dependence on the absolute neutrino mass scale. We fix the scalar field parameters at $\eta_\phi = 0.01$, $\theta = 0$, and $m_\phi = 2\times10^{-24}$~eV.  We note that this parameter set is not expected to cause any significant modification of the energy spectrum at JUNO predicted by the standard oscillation contrary to the cases which have been discussed some time ago in Ref.~\cite{Krnjaic:2017zlz} and recently studied in detail in \cite{Delgadillo:2025wxw}.
The scalar mass corresponds to an oscillation period of approximately 65 years, ensuring the slow-oscillation approximation remains valid throughout the experimental timescales considered. Consequently, this prevents any additional distortion of the JUNO energy spectrum, unlike what is shown in Fig. 1 of ~\cite{Delgadillo:2025wxw}.

For the LBL error bars shown in the figure, we back-propagate the most recent experimental uncertainty. This approach is conservative, as the uncertainties have decreased over time; the actual sensitivity of LBL experiments to temporal modulations of the central value would be even weaker than indicated at earlier times. 
For JUNO, we adopt a time-dependent uncertainty that follows the expected precision improvement with accumulated statistics, as projected in Fig. 8 of Ref.~\cite{JUNO:2022mxj}.  

The vertical arrow in Fig.~\ref{fig:best_fit_time_dep} indicates the critical shift $\delta m^2_{\rm LJ} \sim 4 \times 10^{-5}$ eV$^2$  necessary for a combined analysis of JUNO and LBL data to yield an incorrect conclusion about the mass ordering. Notably, even if nature has chosen normal ordering, the presence of ultralight scalar field with appropriate parameters can lead to an inference of inverted ordering when combining the two measurements. The viability of this confusion scenario depends critically on the neutrino mass: for larger values of $m_3$, the effect is more pronounced. Moreover, the effect would clearly go unnoticed by LBL experiments due to their limited precision, and it may take several years before JUNO can independently identify the temporal trend and resolve the ambiguity, with the timeline ranging from two to more than four years depending on the true value of $m_3$.

\begin{figure*}[t]
    \centering
    \includegraphics[width=0.75\textwidth]{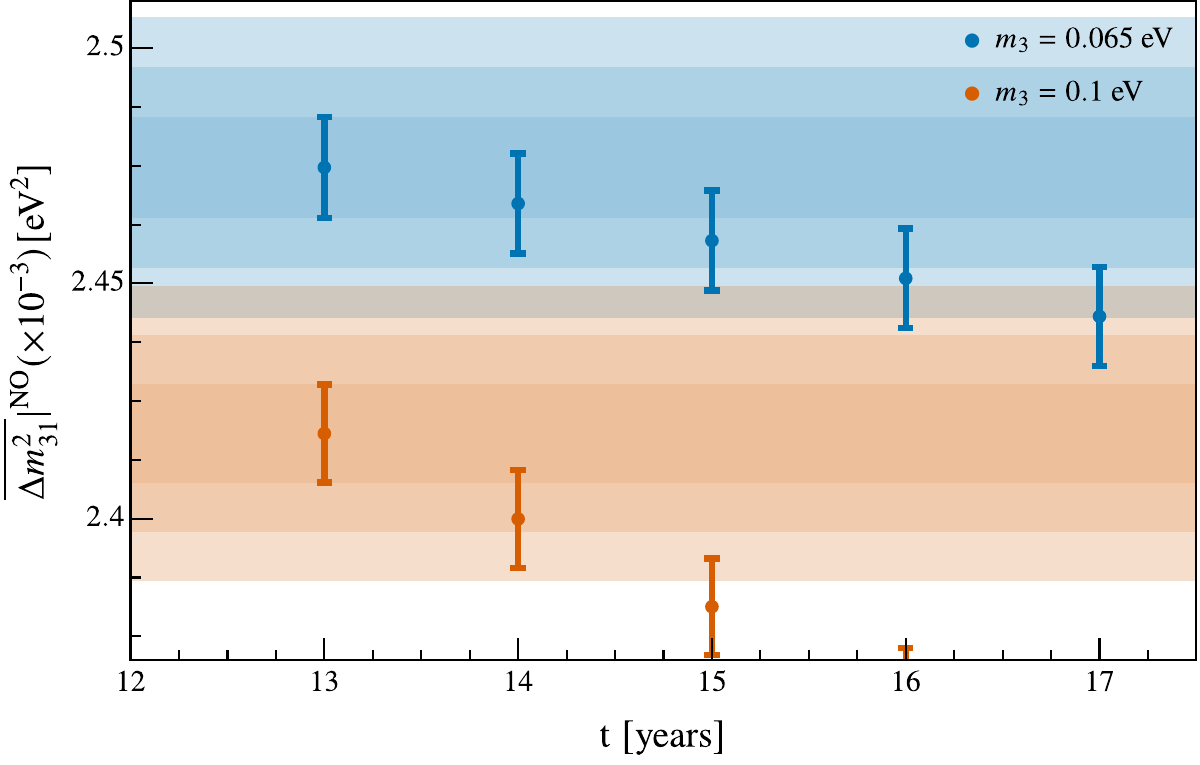}
    \caption{Expected time evolution of JUNO's best-fit atmospheric mass splitting when analyzing data in consecutive one-year slices
    where it is assumed that
    JUNO started taking data at $t=12$ years
    (see text for details), shown for two neutrino mass values. The scalar parameters are fixed as in Fig.~\ref{fig:best_fit_time_dep}. The uncertainty bars are fixed at the one-year precision level. The colored bands represent $1\sigma$ (innermost), $2\sigma$ (middle), and $3\sigma$ (outermost) confidence regions around the initial best-fit value. The systematic drift of the central values demonstrates how time-sliced analysis can reveal the presence of ultralight scalar background. A nearly $3\sigma$ deviation emerges after five years for $m_3 = 0.065$~eV and after only three years for $m_3 = 0.1$~eV.}
    \label{fig:JUNO_1_yr}
\end{figure*}

Given the potential for misidentification of the mass ordering, it would be prudent to also analyze JUNO data in time slices  as well as by combining all data. Fig.~\ref{fig:JUNO_1_yr} illustrates the expected time evolution of JUNO's best-fit value if the experiment reports results using only one-year integration periods. We fix the uncertainty bar at the initial one-year precision level and hold it constant across all time slices. The colored bands indicate the $1\sigma$, $2\sigma$, and $3\sigma$ confidence regions around the initial best-fit value.

The time-sliced analysis reveals how quickly the ultralight scalar scenario could be distinguished from a constant mass splitting. For $m_3 = 0.065$~eV, a $3\sigma$ discrepancy from the initial best fit emerges after approximately five years of operation, while for $m_3 = 0.1$~eV, only two years are required to establish such a deviation.

It is important to emphasize that the presence of the scalar does not impact JUNO's primary objective of determining the mass ordering. When analyzing JUNO data alone, the mass ordering measurement compares oscillation patterns for normal and inverted hierarchies using simultaneously acquired data. Since the scalar-induced modulation affects both orderings identically at any given time, the relative sensitivity between them is preserved. Thus, JUNO can achieve its mass ordering determination while simultaneously probing ultralight scalar parameter.

For a scalar mass $m_\phi = 2 \times 10^{-24}$~eV, recent constraints using the Hubble Space Telescope ultraviolet luminosity functions indicate this field could constitute at most 5\% of the Universe's dark matter~\cite{Lazare:2024uvj}. Additionally, Big Bang Nucleosynthesis bounds constrain the Yukawa coupling to $y_\nu \lesssim 1.8 \times 10^{-11} \sqrt{m_\phi/{\rm eV}}$ for $m_\phi \lesssim 3 \times 10^{-20}$ eV~\cite{Bertolez-Martinez:2025pgm}, based on analysis over the mass range $m_\phi \in (10^{-22}, 10^{-17})$ eV.

Respecting the constraint on $\rho_\phi$ and fixing the neutrino mass to our benchmark value $m_3 = 0.065$~eV, we find
\begin{widetext}
    \begin{equation}
        \eta_\phi \simeq  0.01 \left(\frac{0.065~\text{eV}}{m_3}\right)\left(\frac{\rho_\phi}{0.02\, \rho_{\text{DM}} }\right)^{1/2}\left(\frac{y_\nu}{4 \times 10^{-24} }\right)\left(\frac{2\times 10^{-24}~\text{eV}}{m_\phi }\right),
    \end{equation}
\end{widetext}
where $\rho_{\text{DM}} = 0.4$ GeV/cm$^3$ is the local dark matter density. This suggests JUNO could probe unexplored parameter space independently, with sensitivity depending on the absolute neutrino mass scale.
Finally, we note that future long-baseline experiments, such as Hyper-Kamiokande~\cite{Hyper-Kamiokande:2018ofw} and DUNE~\cite{DUNE:2020jqi},
will be capable of independently identifying
the effect of a time-varying atmospheric
mass-squared difference~\cite{Losada:2021bxx}
 within a few years of operation, due to their significantly higher precision compared to T2K and NOvA.

\section{Conclusions}
\label{sec:conc}

The determination of the neutrino mass ordering, one of the central goals of neutrino oscillation experiments today, enters a decisive new phase with the start of data taking at the JUNO reactor neutrino oscillation experiment in August 2025. JUNO was expressly designed to achieve this objective with unprecedented sensitivity.

A well-known and powerful synergy arises from the combination of high-precision measurements of the effective atmospheric mass-squared splitting obtained from electron antineutrino disappearance at reactor experiments and from muon (anti)neutrino disappearance at long-baseline facilities~\cite{Nunokawa:2005nx}. Fully exploiting this complementarity requires percent-level precision in the $\nu_e$ disappearance effective atmospheric mass splitting, a benchmark that JUNO is expected to reach within only a few months of operation. This prospect has motivated the formulation of a mass ordering sum rule for neutrino disappearance channels~\cite{Parke:2024xre}, demonstrating that by combining data from T2K and NOvA with one year of JUNO operation, the neutrino mass ordering may be established at the $3\sigma$ confidence level, depending on the true value of $\vert \Delta m^2_{31}\vert$.

In this timely context, it is imperative to assess the robustness of this sum rule in the presence of physics beyond the Standard Model. This is the central purpose of our paper. We have identified the conditions under which new physics can impact the sum rule and shown that, in specific scenarios, it may even lead to an incorrect inference of the neutrino mass ordering. As illustrative examples, we examined SNSI~\cite{Ge:2018uhz} and neutrinos coupled to an ultralight scalar field~\cite{Krnjaic:2017zlz, Dev:2020kgz, Berlin:2016woy}. We find that for SNSI, current experimental bounds suppress any observable modification of the sum rule to negligible levels, while in the ultralight scalar case, the interpretation of the ordering requires particular care.

We demonstrate that the ultralight scalar modulation induces a time dependence in JUNO's best-fit determination of the mass-squared splitting governed by two key parameters: the neutrino mass $m_3$ and the modulation strength $\eta_\phi$. If left unrecognized, this effect may ultimately result in an incorrect determination of the neutrino mass ordering when the sum rule is applied, though JUNO's independent mass ordering measurement remains unaffected.

Fortunately, this potential misinterpretation can be independently tested and verified by JUNO itself. Owing to JUNO's unprecedented sub-percent precision measurements, even small temporal variations could become observable. This motivates our strong recommendation that JUNO perform time-sliced analyses of its data to test whether the extracted best-fit value exhibits any systematic drift over time.

More broadly, the mass ordering sum rule emerges not only as a powerful and timely tool for determining the neutrino mass ordering in the JUNO era, but also as a sensitive probe of new physics in the neutrino sector.

\section*{Acknowledgements} 

We are grateful for helpful discussions with Pedro Machado and Stephen Parke and also Pedro Ochoa Ricoux, Liang Zhan and Vanessa Cerrone for useful comments on the earlier version of our manuscript. H.N. thanks Omar Miranda and Luis Delgadillo for useful discussions on the ultralight scalar field scenario. G.F.S.A. is supported by Fermi Forward Discovery Group, LLC under Contracts No. 89243024CSC000002 and DE-SC0010143 with the U.S. Department of Energy, Office of Science, Office of High Energy Physics. R.Z.F. was partially supported by FAPESP under contract No. 2019/04837-9, and by  Conselho Nacional de Desenvolvimento Científico e Tecnológico (CNPq). H.N. is supported by CNPq and CAPES.

\appendix

\section{Varying the scalar phase}
\label{app:scalar_phase}

Throughout the main text, we set the scalar field phase $\theta = 0$ for simplicity,  despite its true value is a priori unknown, as it was enough to illustrate the potential impact of an ultralight scalar coupled to neutrinos when comparing LBL and JUNO data. Here we comment on how different phases may affect the main results.

\begin{figure*}[t]
    \centering
    \includegraphics[width=0.75\textwidth]{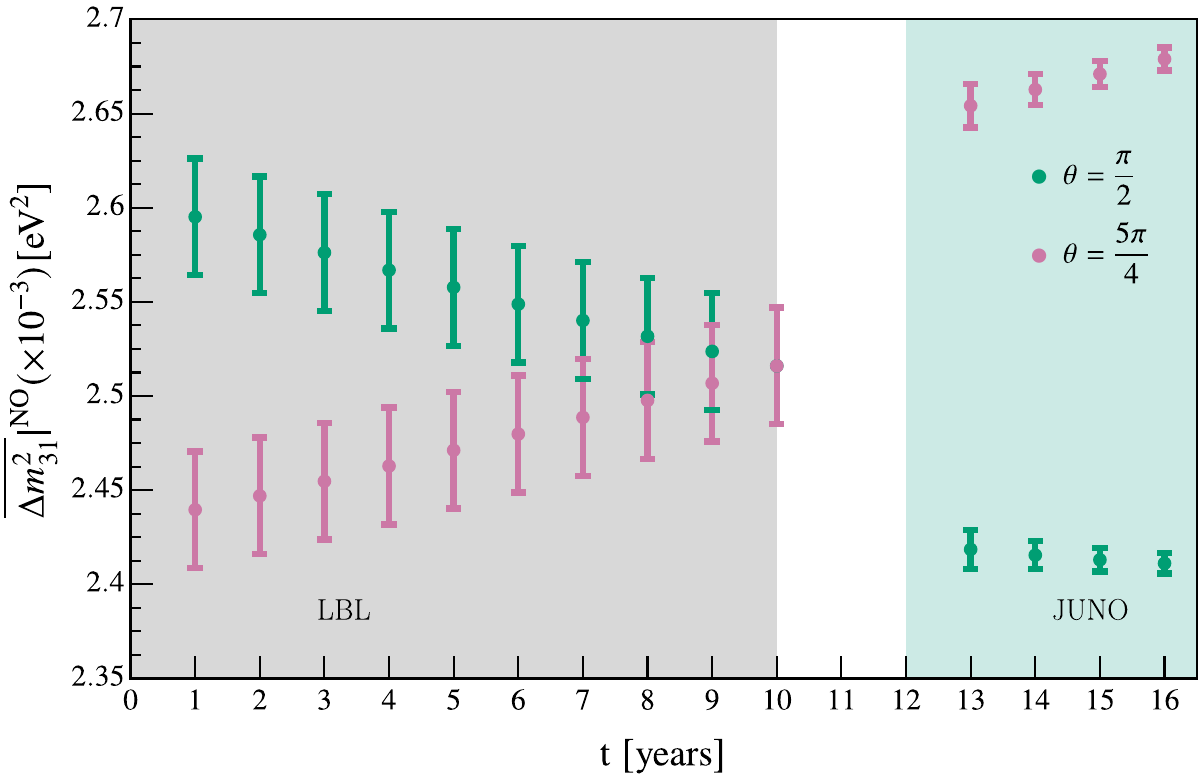}
    \caption{Same as Fig.~\ref{fig:best_fit_time_dep} but for different scalar phases. The neutrino mass is kept fixed at $m_3 = 0.1$~eV and the other scalar parameters are fixed as in Fig.~\ref{fig:best_fit_time_dep}. The phases $\theta = \pi/2$ and $\theta = 5\pi/4$ were selected to maximize the deviation from the $\theta = 0$ case.}
    \label{fig:Scalar phase}
\end{figure*}

The phase modifies the time-averaged mass splitting measured by each experiment. Including $\theta$, the time-averaged atmospheric mass splitting for the LBL experiments becomes
\begin{widetext}
\begin{equation}
    \overline{\Delta m_{31}^2}\big|_{\rm LBL}^{\rm NO} = \Delta m_{31}^2\big|^{\rm NO}_{\rm LBL} + 2m_3^2\eta_\phi\left[\frac{\sin(m_\phi \Delta t_{\rm LBL}+\theta)-\sin\theta}{m_\phi \Delta t_{\rm LBL}} - \frac{\sin(m_\phi \Delta t_{10}+\theta)-\sin\theta}{m_\phi \Delta t_{10}}\right],
\end{equation}
\end{widetext}
while for JUNO,
\begin{widetext}
\begin{equation}
    \overline{\Delta m_{31}^2}\big|_{\rm JU}^{\rm NO} = \Delta m_{31}^2\big|^{\rm NO}_{\rm LBL} - 2m_3^2\eta_\phi\frac{\sin(m_\phi \Delta t_{10}+\theta)-\sin\theta}{m_\phi \Delta t_{10}} + 2m_3^2\eta_\phi\frac{\sin(m_\phi(t_{\rm off}+\Delta t_{\rm JU})+\theta)-\sin(m_\phi t_{\rm off}+\theta)}{m_\phi \Delta t_{\rm JU}}.
\end{equation}
\end{widetext}
The effect of varying $\theta$ is illustrated in Fig.~\ref{fig:Scalar phase}, we fix the neutrino mass to $m_3 = 0.1$~eV, such that the amplitude of the effect is larger and display the results for $\theta = \pi/2$ and $\theta = 5\pi/4$, as they produce the largest departure from $\theta = 0$. We note that the magnitude and sign of the shift $\delta m^2_{\rm LJ}$ between the two experiments depends on this phase. 
Consequently, JUNO is sensitive to both the time modulation and the initial phase of the scalar field. We must note, however, that while Fig.~\ref{fig:Scalar phase} suggests potential phase sensitivity for LBL experiments, the actual precision during the initial years of data collection was significantly lower than illustrated, as the figure assumes the current 
precision for all data points.

\bibliographystyle{apsrev4-1}
\bibliography{main}
  
\end{document}